\documentstyle[eqsecnum,aps]{revtex}

\input{epsf.tex}
\begin{document}
\draft
\twocolumn[\hsize\textwidth\columnwidth\hsize\csname
@twocolumnfalse\endcsname
\preprint{HEP/123-qed}
\renewcommand{\thefootnote}{\alph{footnote}} 
\title{Noise-Induced Phase Space Transport in 
Two-Dimensional Hamiltonian Systems}
\author{Ilya V. Pogorelov\footnote{Electronic address: ilya@phys.ufl.edu}}
\address{Department of Physics, University of Florida, Gainesville, 
Florida 32611\\}
\author{Henry E. Kandrup\footnote{Electronic address: kandrup@astro.ufl.edu}}
\address{ Departent of Astronomy, Department of Physics, and
Institute for Fundamental Theory
\\
University of Florida, Gainesville, Florida 32611}
\date{\today}
\maketitle
\begin{abstract}
First passage time experiments were used to explore the effects of low 
amplitude noise as a source of accelerated phase space diffusion in 
two-dimensional Hamiltonian systems, and these effects were then compared
with the effects of periodic driving. The objective was to quantify and
understand the manner in which ``sticky'' chaotic orbits that, in the absence 
of perturbations, are confined near regular islands for very long times, can 
become ``unstuck'' much more quickly when subjected to even very weak 
perturbations. For both noise and periodic driving, the typical escape time 
scales logarithmically with the amplitude of the perturbation. For 
white noise, the details seem unimportant: Additive and multiplicative noise 
typically have very similar effects, and the presence or absence of a
friction related to the noise by a Fluctuation-Dissipation Theorem is also 
largely irrelevant. Allowing for colored noise {\it can} significantly 
decrease the efficacy of the perturbation, but only when the autocorrelation 
time, which vanishes for white noise, becomes so large
that there is little power at frequencies comparable to 
the natural frequencies of the unperturbed orbit. Similarly, periodic driving
is relatively inefficient when the driving frequency is not comparable to 
these natural frequencies. This suggests strongly that noise-induced extrinsic
diffusion, like modulational diffusion associated with periodic driving, is a 
resonance phenomenon. The logarithmic dependence of the escape time on 
amplitude reflects the fact that the time required for perturbed and 
unperturbed orbits to diverge a given distance scales logarithmically in the 
amplitude of the perturbation.
\end{abstract}
]
\narrowtext

\section{MOTIVATION}
\label{sec:level1}
It is well known that a complex phase space containing large measures of both
regular and chaotic orbits is often partitioned by such partial obstructions
as cantori \cite{1} or Arnold webs \cite{2} which, although not serving as 
absolute barriers, can significantly impede the motion of a chaotic 
orbit through a connected phase space region. Indeed, the fact that, in 
two-dimensional Hamiltonian systems, chaotic orbits can be ``stuck'' near 
regular islands for very long times was discovered empirically \cite{3}  
long before the existence of cantori was proven \cite{4}.

It has also been long known that low amplitude stochastic perturbations can
accelerate Hamiltonian phase space transport by enabling orbits to traverse 
these partial barriers. This was, e.g., explored by Lieberman and Lichtenberg 
\cite{5}, who investigated how motion described by the simplified Ulam version 
of the Fermi acceleration map \cite{6} is impacted by random perturbations,
allowing for the modified equations \cite{7}
\begin{displaymath}
\hskip -1.1in u_{n+1}=|u_{n}+{\psi}_{n}-{1\over 2}\,|, 
\end{displaymath}
\begin{equation}
{\psi}_{n+1}={\psi}_{n}+{M\over u_{n+1}} + {\Delta}{\psi}, \qquad
{\rm mod}\; 1 ,
\end{equation}
where the ``noise'' ${\Delta}{\psi}$ corresponds to a random phase shift 
uniformly sampling an interval $[-{\varphi},+{\varphi}]$. 

That stochastic perturbations can have such effects on Hamiltonian 
systems is important in understanding the limitations of simple models of
real systems. In the absence of all ``perturbations'' and any other
irregularities, the chaotic phase space associated with some idealised two- or 
three-dimensional Hamiltonian system may be partitioned into regions which are 
effectively distinct over relatively short time scales. However, even very 
weak perturbations of the idealised model, so small as to seem irrelevant on 
dimensional grounds, can blur these barriers and permit a single orbit to move 
from one region to another on surprisingly short time scales.

One practical setting where this may be important is in understanding how, in 
the context of the core-halo model \cite{8} of mismatched
charged particle beams, the focusing of an accelerator beam can be corrupted 
by imperfections in the magnetic fields. To the extent that such
irregularities can be modeled as noise, there is the concern that noise-induced
diffusion can result in particles in the beam becoming sufficiently defocused
as to hit the walls of the container, which is a disaster. Work in
this area is currently focused on obtaining realistic estimates of the noise
amplitude and the form of the power spectrum \cite{9}.

Another setting is in galactic astronomy. Recent observations
indicating (i) that many/most galaxies are genuinely triaxial, i.e., neither 
spherical nor axisymmetric, and (ii) that they contain a pronounced central 
mass concentration suggest strongly that the self-consistently determined bulk 
gravitational potential associated with a galaxy contains significant measures 
of both regular and chaotic orbits \cite{10}. It was originally expected that, 
in such complex potentials, regular orbits would provide the skeleton to 
support the triaxial structure, and that chaotic orbits would serve to fill in 
the remaining flesh of the self-consistent equilibrium \cite{11}. However, it 
appears that, in many cases, much of the expected role of regular orbits 
must be played by ``sticky'' chaotic orbit segments since, as a result of 
resonance overlap, the measure of regular orbits in certain critical regions 
is very small, albeit nonzero \cite{12}. The obvious question is: can low 
amplitude perturbations reflecting internal substructures like gas clouds and 
individual stars or the effects of an external environment destabilise a 
near-equilibrium on a time scale short compared with the age of the 
Universe? Preliminary work would suggest that they can \cite{13}.

In both these settings, one knows that weak perturbations will eventually 
trigger significant changes in energy on some fiducial relaxation time $t_{R}$,
which implies that they could have a significant effect. This, however, is
not the critical issue here. Rather, the question is whether low amplitude
perturbations can have significant effects already on a time scale short 
compared with the time scale on which the value of the energy, or any other
isolating integral, changes significantly. 

In understanding the potential effects of such low amplitude ``noise,'' there
are at least three important questions which need to be addressed:
\par\noindent
1. {\it How does the effect depend on the amplitude of the noise?} Is there
a threshhold amplitude below which the noise is essentially irrelevant, or
do the effects turn on more gradually? Does the efficacy of the perturbation
scale as a simple power of the amplitude or does one see something more 
subtle? 
\par\noindent
2. {\it To what extent do the details of the noise matter?} For some problems,
such as energy barrier penetration, additive (i.e., state-independent) and
multiplicative (i.e., state-dependent) noises can yield very different results 
\cite{14}. However, the physics here is not the same since
one is not dealing with a barrier which, in the absence of perturbations, is
absolute. Rather, one is dealing with an {\it entropy barrier} \cite{15}. It 
would seem that the problem of diffusion through cantori or along an Arnold 
web is more similar to problems involving chaotic scattering \cite{16} or 
escapes of unbound orbits from a complex Hamiltonian system \cite{17} where, 
in the absence of perturbations, the requisite escape channels exist and it 
is only a matter of how fast any given orbit can find one.
\par\noindent
3. {\it Why does noise lead to accelerated phase space transport?} Granted
that the physics is different from diffusion in energy, what is the correct
physics? One possibility is that introducing noise simply fuzzes out the 
details of a purely Hamiltonian evolution that are ensured by Liouville's
Theorem, thus enabling orbits to breach gaps which would otherwise
be impenetrable. However, something very different might be responsible for
what is seen.

This paper aims to address these questions for two-dimensional Hamiltonian
systems by performing {\it first passage time experiments.} What this entails
is identifying chaotic orbits which, in the absence of any perturbations, 
remain ``stuck'' near regular islands for very long times, and determining
how the introduction of weak noise reduces the escape time. The experiments
that were performed and interpreted involved both additive and multiplicative 
noise. They also allowed for both white noise, which is delta-correlated in 
time and has a flat power spectrum, and colored noise, which has a finite 
autocorrelation time, so that the power spectrum effectively cuts off for 
large frequencies. Finally, the
experiments allowed for both external noise, presumed to exist in and of
itself, and internal noise, which is accompanied by a friction that
is related to the noise by a Fluctuation-Dissipation Theorem \cite{18}. To gain
additional insights, the results of these noisy experiments were also compared 
with experiments in which the unperturbed initial conditions were evolved in 
the presence of low amplitude periodic driving, so that the breaching of 
cantori could be triggered by modulational diffusion \cite{7}.

Section II describes the experiments that were performed, and the following 
three sections  report the results. Section III summarises the effects of low 
amplitude periodic driving, indicating the relative importance of the 
amplitude and frequency of the perturbation. Section IV describes the effects 
of different sorts of white noises; and Section V generalises the preceding
section to the case of colored noise. Section VI concludes by summarising
the evidence that, like periodic driving, noise-induced extrinsic diffusion
through cantori is a resonance phenomemon which requires substantial power at 
frequencies comparable to the natural frequencies of the unperturbed orbit, 
and which has an efficacy that scales logarthmically in the amplitude of the 
perturbation.

\section{A DESCRIPTION OF THE COMPUTATIONS} 

The experiments described here were performed for orbits evolved in two 
representative two-dimensional potentials, namely the so-called dihedral 
potential \cite{19} for one particular set of parameter values, for which the 
Hamiltonian takes the form
\begin{equation}
H={1\over 2}{\Bigl(}p_{x}^{2}+p_{y}^{2}{\Bigr)}
-(x^{2}+y^{2})+{1\over 4} (x^{2}+y^{2})^{2}-{1\over 4}x^{2}y^{2},
\end{equation}
and the sixth order truncation of the Toda lattice potential \cite{20}, for 
which
\begin{displaymath}
H={1\over 2}{\Bigl(}p_{x}^{2}+p_{y}^{2}{\Bigr)}+ {1\over
2}{\Bigl(}x^{2}+y^{2}{\Bigr)}+x^{2}y-{1\over 3}y^{3} +
\end{displaymath}
\begin{displaymath}
{1\over 2}x^{4}+
x^{2}y^{2}+{1\over 2}y^{4}+x^{4}y+{2\over
3}x^{2}y^{3}-{1\over 3}y^{5} +
\end{displaymath}
\begin{equation}
{1\over 5}x^{6}+x^{4}y^{2}+{1\over3}x^{2}y^{4}+{11\over 45}y^{6}.
\end{equation}
Extensive explorations of orbits in these Hamiltonians would suggest that, in
many respects, these potentials are generic in the set of nonintegrable 
potentials admitting global stochasticity. This is consistent with the fact 
that the experiments
performed for this paper yielded similar results both potentials. However,
these potentials {\it are} special in the sense that they admit discrete
symmetries: the dihedral potential is invariant under a rotation by ${\pi}/4$;
the truncated Toda potential is invariant under a rotation by $2{\pi}/3$. It 
should
be noted for future reference that, for relatively low energies, $E<40$ or so,
a characteristic orbital time scale in each potential corresponds to a time
$t{\;}{\sim}{\;}1-3$, so that most of the power in typical orbits is in 
frequencies ${\omega}{\;}{\sim}{\;}1-5$. 

In both potentials it is easy to find ``sticky'' \cite{3} chaotic orbits which,
visually, are very nearly indistinguishable from regular orbits for 
comparatively long times (although they have short time Lyapunov exponents 
sufficiently large that they must be chaotic). Three examples are exhibited
in the left hand panels of FIG. 1, namely two orbits in the 
dihedral potential, with energies $E=10$ and $E=20$, and an orbit in the 
truncated Toda potential with $E=20$. The orbit in FIG.1 (a) resembles 
closely what a galactic astronomer would term a 
regular loop orbit; the orbits in FIG. 1 (d) resembles a regular fish. The
orbit in FIG. 1 (g) is less familiar, but would again seem nearly regular.
The important point, then, is that if the orbit is integrated for a somewhat
longer interval, its behaviour exhibits an abrupt qualitative change. This
is illustrated in the center panels of FIG. 1, which exhibit the same initial 
conditions, each integrated for an interval twice as long. The first two 
orbits are no longer centrophobic, and the third has so changed 
as to manifest explicitly the discrete $2{\pi}/3$ rotation symmetry of the 
truncated Toda potential. In each case, the orbit is far more chaotic, as is 
readily confirmed by the computation of a Lyapunov exponent.

The transition from nearly regular, ``sticky'' behaviour to something 
more manifestly chaotic occurs once the orbit has diffused through one or 
more cantori that surround a regular phase space island \cite{1}\cite{4}. The 
orbit begins
chaotic and remains chaotic throughout, but its basic properties exhibit 
significant qualitative changes after the orbit has escaped through the cantori
to become ``unconfined.'' The precise objective of the work described here
is to determine how the time required for chaotic orbits to change from 
sticky to unconfined is altered when the orbit is perturbed by low amplitude 
perturbations. 

Determining the precise location of the outermost cantorus is possible, albeit 
exceedingly tedious \cite{21}. Fortunately, however, this is not 
essential to estimate with reasonable accuracy when a ``sticky'' 
orbit has become ``unstuck.'' Once the orbit has breached the outermost 
confining cantorus, it will typically move quickly to probe large portions of 
the accessible configuration space regions which were inaccessible before this 
escape. Moreover, escape is accompanied by an abrupt increase in the value of 
the largest short time Lyapunov exponent \cite{22}, this reflecting the fact 
that ``sticky'' chaotic orbit segments confined near regular islands tend to 
be less unstable than unconfined chaotic segments far from any 
regular island \cite{23}.

As a practical matter, the {\it first escape time} for a  sticky chaotic
orbit was identified by (1) using simple polynomial formulae to delineate
approximately the configuration space region to which the orbit is originally
confined, and then (2) determining the first time that, with or without 
perturbations, the orbit leaves this special region. To check that the 
escape criterion was reasonable, two tests were performed: It was verified
that, with or without perturbations, small changes in the precise definition 
have only minimal effects on the computed first escape time; and that, for the 
case of unperturbed orbits, the time of escape corresponds to a time when the 
largest short time Lyapunov exponent exhibits an abrupt increase.

This prescription allowed one to identify with reasonable accuracy transitions
from sticky chaotic to unconfined chaotic behaviour, but {\it not} from 
chaotic to regular. The constant energy surface contains {\it KAM} tori, which
serve as absolute boundaries between regular and chaotic behaviour, so that
an unperturbed orbit that starts as chaotic can never become regular. If, 
however, the orbit is perturbed, the energy is no longer conserved, and it
becomes possible in some cases for an initially chaotic orbit to become
regular.

The experiments described in this paper involved generating ensembles of 
perturbed orbits and then extracting statistical properties from these
ensembles. In this setting, two different diagnostics proved especially
useful:
\par\noindent
1. {\it The time $T(0.01)$ required for one percent of the orbits in the 
ensemble to escape.} As described in the following Sections, escapes do not 
begin immediately. Rather, there is typically a relatively extended initial
period, the duration of which depended on the form of the perturbation, during
which no escapes are observed. Perhaps the most obvious number to record 
would be time when the first orbit in the ensemble escaped. However, it was
found that, in a nonnegligible fraction of the experiments that were performed
-- perhaps 5-10\% -- one orbit often escapes long before any of the others. 
For this reason, it seemed more reasonable to track a diagnostic that is less
sensitive to comparatively rare exceptions. 
\par\noindent
2. {\it The initial escape rate ${\Lambda}$}. In many, albeit not all, cases 
it was found that, once the escape
process ``turns on'' at (say) time $t_{0}$, orbits escape in a fashion which, 
at least initially, is consistent with a Poisson process, with $N(t)$, the 
fraction of the orbits which have not escaped, decreasing exponentially:
\begin{equation}
N(t)=N_{0}\exp [-{\Lambda}(t-t_{0})] .
\end{equation}

The experiments with periodic driving involved solving an evolution equation of
the form
\begin{equation}
{d^{2}{\bf r}\over dt^{2}}=-{\nabla}V({\bf r})+A\sin ({\omega}t+{\varphi})
\hat{{\bf r}}.
\end{equation}
The driving was thus characterised by three parameters, namely the 
frequency ${\omega}$, the amplitude $A$, and the phase ${\varphi}$. Usually
but not always the phase ${\varphi}$ was set equal to zero. Ensembles of
periodically driven orbits were generated by (1) specifying a frequency
interval $[{\omega},{\omega}+{\Delta}]$, (2) sampling this 
interval uniformly to select a collection of (usually) 1000 driving 
frequencies, and then (3) integrating the same initial condition with the
same amplitude $A$ for each of these frequencies. When looking at relatively
low frequencies, $0<{\omega}<100$, the frequency range was taken to be 
${\Delta}=1.0$. For higher frequencies, $100<{\omega}<1000$, the range
${\Delta}=10.0$.

The experiments involving intrinsic noise entailed solving Langevin equations 
of the form
\begin{equation}
{d^{2}{\bf r}\over dt^{2}}=-{\nabla}V({\bf r})-{\eta}{\bf v}+
{\bf F}, 
\end{equation}
with ${\eta}={\eta}({\bf v})$ and ${\bf F}$ homogeneous Gaussian noise 
characterised by its first two moments:
\begin{displaymath}
{\langle}F_{a}(t){\rangle}=0 \qquad {\rm and}
\end{displaymath}
\begin{equation}
{\langle}F_{a}(t_{1})F_{b}(t_{2}){\rangle}=
{\delta}_{ab}\;K({\bf v},t_{1}-t_{2}), 
\hskip .2in (a,b=x,y).
\end{equation}
$K({\bf v},{\tau})$ is the autocorrelation function. For the case of 
delta-correlated white noise, 
\begin{equation}
K({\bf v},{\tau})=2{\Theta}{\eta}({\bf v}){\delta}_{D}({\tau}), 
\end{equation}
where ${\Theta}$ denotes a characteristic temperature, the friction and noise
being related by a Fluctuation-Dissipation Theorem. Experiments involving
extrinsic noise proceeded identically, except that the friction was turned
off, so that
\begin{equation}
{d^{2}{\bf r}\over dt^{2}}=-{\nabla}V({\bf r})+{\bf F}. 
\end{equation}

White noise simulations were performed using an algorithm developed by 
Griner et al \cite{24} (see also \cite{25}). Colored noise simulations were 
performed using a more complex algorithm described in Section V. The 
experiments with delta-correlated white noise allowed both for additive noise,
where ${\eta}$ is a constant, and multiplicative noise, where ${\eta}$
is a nontrivial function of ${\bf v}$. The experiments with 
colored noise involved two different choices for the form of 
$K({\tau})$, in each case allowing for a parameter ${\alpha}$ which 
characterised the temporal width of the autocorrelation function. In every
case, ensembles of orbits with the same initial conditions were generated by
freezing the form and amplitude of the noise and performing multiple 
realisations of the same random process using different pseudo-random seeds.

\section{PERIODIC DRIVING AND MODULATIONAL DIFFUSION}

Experiments involving multiple integrations of the same initial condition
reveal that escape is (at least) a two-stage process: In general there is an 
initial interval, often quite extended, during which no escapes occur. Only 
after this interval is there an abrupt onset of escapes which, at least for 
relatively early times, can be well modeled as a Poisson process, where $N(t)$,
the fraction of the orbits that
have not yet escaped, decreases exponentially. An example of this behaviour
is illustrated in FIG. 2, which was generated for the initial condition 
exhibited in FIG 1 (a), allowing for a frequency interval $2.0{\;}{\le}{\;}
{\omega}{\;}{\le}{\;}3.0$ and an amplitude $A=10^{-2.5}$. The straight line
exhibits a linear fit to the interval $T(0.01){\;}{\le}{\;}t{\;}{\le}{\;}300$.

As asserted already, at time only somewhat larger than $T(0.01)$, $N(t)$ 
appears to decrease exponentially. However, for $t>400$ or so it is clear
from FIG. 2 that $N(t)$ decreases more slowly. One plausible 
interpretation of this later subexponential decay is 
that some of the initially sticky chaotic orbits have becomes trapped even
closer to the regular island or, in some cases, have actually become regular,
so that escape becomes much more difficult if not impossible. This 
interpretation was tested by turning off the periodic driving at a
late time $t=1024$ and computing both the orbit and an approximation to the
largest short time Lyapunov exponent for the interval $1024<t<3072$. An
analysis of the resulting output indicated that, in some cases, the orbits
which had not escaped by $t=1024$ had in fact become regular.

For fixed frequency interval and phase, both $T(0.01)$, a measure of the 
time before escapes begin, and ${\Lambda}$, the initial escape rate once 
escapes have begun, typically scale logarithmically in $A$, the amplitude of 
the driving. Six examples of this behaviour are provided in FIG. 3, these
corresponding to the three initial conditions exhibited in FIG. 1 
for two different frequency intervals, namely $0.0{\;}{\le}{\;}{\omega}
{\;}{\le}{\;}1.0$ and $2.0{\;}{\le}{\;}{\omega}{\;}{\le}{\;}3.0$. In each case,
the size of the error bar has been set equal the difference between $T(0.01)$ 
and the time at which
the first orbit escapes. In most cases, this difference is small, but in some
cases it becomes appreciable. The fact that the curve is not exactly linear,
and that it levels out for certain ranges of amplitude, is not an obvious
finite number effect. Doubling the number of frequencies that were sampled, 
and hence the number of orbits, does not significantly impact the overall 
smoothness of the curve.

Before the onset of escapes, the {\it rms} deviation ${\delta}r_{rms}$ between
perturbed and unperturbed orbits typically grows as
\begin{equation}
$${\delta}r_{rms}{\;}{\propto}\;A\exp({\chi}t), 
\end{equation}
where $A$ is the driving amplitude and ${\chi}$ is comparable 
to the positive short time Lyapunov exponent for the unperturbed orbit. 
The {\it rms} deviation ${\delta}E_{rms}$ also varies linearly with $A$, but 
exhibits a much weaker time dependence. That both these quantities scale
linearly in $A$ is hardly surprising since periodic driving is a coherent
process. The different time dependences reflect the fact that, although nearby 
chaotic orbits tend to diverge exponentially in configuration space, with or 
without small perturbations, energy is conserved absolutely in the absence of 
perturbations. That $T(0.01)$ scales as $\log A$ means that escapes begin when
${\delta}r_{rms}$, rather than ${\delta}E_{rms}$, assumes a roughly constant 
value, independent of the amplitude $A$. The characteristic value when
escapes begin is typically ${\delta}r_{rms}{\;}{\sim}{\;}1-2$, which implies
that the perturbed orbits have dispersed to probe most of the region inside
the confining cantori.

This leads to a natural interpretation of the escape process: Early on, the
perturbed orbits remain relatively close to the unperturbed orbit, so that
it is unlikely that they will be able to escape. (The initial conditions were
so chosen that, in the absence of perturbations, escape only occurs at a 
comparatively late time!) Eventually, however, the perturbed orbits will have
spread out to sample more or less uniformly some region inside the bounding 
cantori. Once this has happened, orbits will begin to escape ``at random'' in
a fashion that samples a Poisson process. If the holes were
very large, one might expect that the escape rate at this stage would be
nearly independent of amplitude. Given, however, that orbits still have 
to ``hunt'' for tiny escape channels, one might expect that ${\Lambda}$
also depends logarithmically on the amplitude of the perturbation. 

This interpretation is consistent with the expectation that an initially 
localised ensemble of chaotic orbits will exhibit an exponential in time 
approach towards a near-invariant distribution that corresponds to a 
near-uniform population of those accessible phase space regions not obstructed 
by cantori \cite{26}. It is also qualitatively similar to what appears to 
happen when considering the escape of energetically unbound orbits from a 
complicated two-dimensional potential \cite{27}.

Periodic driving tends to yield the smallest $T(0.01)$ and largest ${\Lambda}$ 
for driving frequencies ${\omega}$ comparable to 
the natural frequencies of the unperturbed orbits. For example a plot of 
$T(0.01)$ as a function of ${\omega}$ for fixed amplitude $A$ and phase 
${\varphi}$ typically exhibits the smallest values of $T(0.01)$ for 
${\omega}{\;}{\sim}{\;}1-3$ and an abrupt increase for somewhat larger
frequencies. However, low amplitude driving can still have an appreciable
effect on the time of escape even when the driving frequency is much larger
than the natural frequencies of the unperturbed orbit. For example, $T(0.01)$
can be significantly shorter than the escape time for an unperturbed orbit
even for driving frequencies as large as ${\omega}{\;}{\sim}{\;}1000$.

Three examples of how $T(0.01)$ varies with ${\omega}$ for fixed $A$ and
${\phi}$ are exhibited in FIG. 4. The three left panels plot $T(0.01)$ as a 
function of ${\omega}$ for $0{\;}{\le}{\;}{\omega}{\;}{\le}{\;}40$. 
The three right panels plot $T(0.01)$ as a function of $\log\,{\omega}$ for
$1{\;}{\le}{\;}{\omega}{\;}{\le}{\;}1000$. In some cases $T(0.01)$ varies
smoothly as a function of ${\omega}$ for ${\omega}{\;}{\gg}{\;}10$; in other
cases, considerably more irregularity is evident. In either case, however,
it is apparent that, overall, the efficacy of the driving is set by the 
logarithm of the driving frequency. $T(0.01)$ tends to increase linearly in
$\log\,\omega$. Given the plausible hypothesis that this accelerated escape is
a resonance phenomenon involving a coupling between the driving frequency and
the natural frequencies of the unperturbed orbits, the fact that high 
frequencies still have an appreciable effect can be interpreted as implying 
that, even though the unperturbed orbit has little power at high 
frequencies, periodic driving can couple via higher order harmonics.

When, for fixed $A$ and ${\varphi}$, $T(0.01)$ and ${\Lambda}$ are 
comparatively smooth functions of driving frequency, $T(0.01)$ tends to 
exhibit only a relatively weak dependence on the phase. Different values of 
${\varphi}$ tend to yield comparable escape times. If, alternatively, 
$T(0.01)$ depends sensitively on ${\omega}$, it is more likely that the escape 
time also depends sensitively on ${\varphi}$. However, this trend is not
uniform. In some cases, varying ${\varphi}$ continuously from $0$ to $2{\pi}$ 
changes $T(0.01)$ by no more than 10\%. In other cases, $T(0.01)$ can vary by 
a factor of four, or more. Finally, it should be noted that the importance of
noise in accelerating diffusion through cantori can depend sensitively on the 
details of the orbit. Consider, e.g., two initial conditions in the same 
potential with the same energy which probe nearby phase space regions and 
which, in the absence of perturbations, lead to orbits that escape at 
comparable times. There is no guarantee that ensembles of periodically driven
orbits generated from these different initial conditions and evolved with
the same amplitudes, phases, and driving frequencies will exhibit similar
values of $T(0.01)$ and ${\Lambda}$, even if the unperturbed orbits have power 
spectra that are almost identical.

\section{ WHITE NOISE}
For stationary Gaussian white noise with zero mean, everything is 
characterised by the autocorrelation function
$K(t_{1}-t_{2})=
{\Theta}{\eta}({\bf r},{\bf v}){\delta}_{D}(t_{1}-t_{2}),$
the form of which is determined in turn by ${\eta}$. Choosing ${\eta}$ to be
constant yields additive noise. Allowing for a nontrivial 
dependence on ${\bf r}$ or ${\bf v}$ yields multiplicative noise. One 
aim of the work described here was to determine the extent to which
the detailed form of the noise matters. This was done by first
performing experiments involving additive noise, and then comparing the
results with experiments that involved multiplicative
noise of two forms, namely ${\eta}{\;}{\propto}{\;}v^{2}$ and
${\eta}{\;}{\propto}{\;}v^{-2}$, where $v$ denotes the orbital speed. The
importance of friction was tested by comparing experiments that included 
a friction related to the noise by a Fluctuation-Dissipation Theorem\cite{18} 
with experiments with no friction at all.

If the friction and noise are to mimic internal degrees of freedom
that are ignored in a mean field description, one anticipates on dimensional
grounds that the temperature ${\Theta}$ will be comparable to a typical
orbital energy. For this reason, most of the experiments that were performed, 
including those described here, involved freezing the temperature at a 
value ${\Theta}{\;}{\sim}{\;}E$ and exploring the effects of varying the 
amplitude of ${\eta}$. The relative normalisations of the multiplicative and 
additive noises were fixed by setting 
\begin{equation}
$${\eta}({\bf v})={\eta}_{0}(v/{\langle}v{\rangle})^{\pm 2}, 
\end{equation}
where ${\eta}_{0}$ denotes the constant ${\eta}$ appropriate for additive
noise and ${\langle}v{\rangle}$ denotes the average speed of the unperturbed
orbit. Comparing additive and multiplicative noise entailed comparing 
experiments with the same ${\eta}_{0}$.

Considering only two forms of multiplicative noise involves probing
the tip of an iceberg: other multiplicative noises
could in principle have very different effects. However, the two cases 
examined here do allow one to ask whether the overall effect of the 
friction and noise can change significantly if one allows the statistics
of the noise to vary along an orbit. The particular forms chosen here were
motivated by two considerations: (1) If the noise is intended to mimic 
discreteness effects in a plasma or a galaxy (i.e., electrostatic or 
gravitational Rutherford scattering), the friction should depend on velocity
\cite{28}. (2) Allowing for a relatively strong dependence on speed, 
${\eta}{\;}{\propto}{\;}
v^{\pm 2}$, should make even relatively small differences comparatively easy
to see.

Overall, the effects of white noise are very similar to the effects of
periodic driving. In particular, escape was again observed to be a two stage
process, involving an initial interval during which different realisations
of the same initial condition diverge inside the confining cantori, followed
by an abrupt onset of escapes which, at least initially, is well approximated
as a Poisson process. Moreover, as for the case of periodic driving, $N(t)$
decreases subexponentially at late times, possibly because some
of the noisy orbits have become regular or, at least, more closely trapped 
near a regular island.

FIGURE 5 exhibits plots of $\log N(t)$ generated for one representative initial
condition, corresponding to the orbit 
in FIG. 1 (a). Here the experiments involved additive noise and 
friction related by a Fluctuation-Dissipation Theorem, with a fixed 
${\Theta}=10$ and $10^{-9}{\;}{\le}{\;}{\eta}_{0}{\;}{\le}{\;}10^{-4}$.
It is evident that, as ${\eta}_{0}$ decreases and the friction and
noise become weaker, the escape time $T(0.01)$ increases and the escape rate
${\Lambda}$ decreases.

More careful examination reveals further that, for fixed ${\Theta}$, both 
$T(0.01)$ and ${\Lambda}$ scale logarithmically in ${\eta}_{0}$. This is 
illustrated in FIG. 6, which exhibits $T(0.01)$ and a best fit value of
${\Lambda}$ for the ensembles used to construct FIG. 5 (along with some 
ensembles with intermediate values of ${\eta}$). This logarithmic
dependence can be understood by analogy with what was observed for 
periodic driving if one notes that, in the presence
of noise, the {\it rms} deviation between perturbed and unperturbed orbits 
typically scales as 
\begin{equation}
$${\delta}r_{rms}{\;}{\propto}{\;}(\eta{\Theta})^{1/2}\exp({\chi}t),
\end{equation}
a conclusion that can be motivated theoretically \cite{29} and and has been 
confirmed computationally \cite{13}.

It is interesting that this two-stage evolution -- an epoch without escapes 
followed by an epoch with escapes apparently sampling a Poisson process -- can
also be observed in the absence of noise if one considers a strongly localised
ensemble of initial conditions trapped near a regular island and ascertains
the time at which each member of the ensemble escapes. For example, an 
ensemble of orbits sampling a cell of size $0.002$ centered about the initial
condition used to generate FIG. 6 yielded $T(0.01)=310$ and 
${\Lambda}=0.000587$, which should be compared with the values $T(0.01)=131$
and ${\Lambda}=0.00153$ resulting for a single initial condition evolved 
with ${\eta}=10^{-9}$.

Perhaps the most significant conclusion about white noise is that, at least
for the examples considered here, the details are largely irrelevant.
For fixed ${\Theta}$ and ${\eta}_{0}$, the values of the escape time 
$T(0.01)$ and the decay rate ${\Lambda}$ are both essentially the same for 
the simulations with additive noise and those with multiplicative noise 
with ${\eta}{\;}{\propto}{\;}v^{\pm 2}$. The computed values of $T(0.01)$ and 
${\Lambda}$ are also nearly independent of whether or not one allows for a
friction term. The only significant differences between simulations with
and without friction arise at late times when the energies of individual orbits
have changed appreciably from their initial values. In this case, allowing
for a friction term to counterbalance the noise assures that, overall, the
energies of the orbits exhibit smaller changes, so that the ensembles evolved
with noise tend to have somewhat smaller changes in energy.

An example of this insensitivity is provided in FIG. 7, which was generated
from orbits with the initial condition of FIG. 1 (a), with ${\Theta}=10$ 
and ${\eta}_{0}=10^{-5}$. The solid, dot-dashed, and triple-dot-dashed curves
exhibit $N(t)$ for ensembles evolved in the presence of both friction and
noise, incorporating, respectively, additive noise, multiplicative noise with
${\eta}_{0}{\;}{\propto}{\;}v^{2}$, and multiplicative noise with
${\eta}_{0}{\;}{\propto}{\;}v^{-2}$. The dashed curve corresponds to an 
ensemble evolved with additive noise in the absence of friction. The obvious 
point is that, for a very long time, these curves are nearly indistinguishable.

This insensitivity is again consistent with the hypothesis that noise-induced
diffusion through cantori is a resonance phenomenon, and that the only 
thing that matters is that the noise have significant power at frequencies
comparable to the natural frequencies of the unperturbed orbit. If, however,
one alters the form of the autocorrelation function so as to suppress power
at frequencies comparable to these natural frequencies, one would expect that
the effects of the noise should decrease, so that $T(0.01)$ increases and 
${\Lambda}$ decreases. The extent to which this is true is discussed in 
Section V.

\section{COLORED NOISE}
The objective of the experiments described here was to explore 
the effects of random perturbations with autocorrelation times 
sufficiently long that they cannot be modeled as delta-correlated white noise. 
Once again it was assumed that the noise is stationary and Gaussian with zero 
mean and, for simplicity, attention was restricted to noise that is 
additive. However, it was no longer assumed that $K({\tau})$ is 
delta-correlated in time. 

Attention focused on two types of colored noise: The first is 
generated by the Ornstein-Uhlenbeck process (see, e.g., \cite{18}), and is 
characterised by an autocorrelation function $K({\tau})$ that decays 
exponentially, i.e.,
\begin{equation}
$$ K({\tau})={\sigma}_{col}^{2}\,\exp(-{\alpha}|{\tau}|).  
\end{equation}
The second involves an exponential decay modulated by power law:
\begin{equation}
$$K(\tau) = {{\sigma}_{col}}^{\hskip -.14 in 2} {\hskip .07 in }
\exp(- \alpha | \tau |) {\Bigl(} 1+ \alpha | \tau | +  \frac{{\alpha}^2}{3} 
{\tau}^2  {\Bigr)}.
\end{equation}
The corresponding spectral densities are, respectively,
\begin{equation}
S({\omega})={{\sigma}_{col}^{2}\over {\pi}}\,
{{\alpha}\over {\omega}^{2}+{\alpha}^{2}} 
\end{equation}
and
\begin{equation}
S(\omega) = \frac{8 {{\sigma}_{col}}^{ \hskip -.14 in  2 }}  
{3 \pi} \frac{{\alpha}^5}{({\omega}^2 +{\alpha}^2)^3}.
\end{equation}
The autocorrelation times are $t_{c}=1/{\alpha}$ and $t_{c}=2/{\alpha}$.
In both cases, white noise corresponds to a singular 
limit with ${\alpha} \to \infty$ and ${\sigma}_{col}^{2} \to\infty$, but 
${\sigma}_{col}^{2}/{\alpha}\to$ const. As for the case of multiplicative 
noise, these two examples only probe the tip of an iceberg.  However, 
an analysis of their effects 
{\it does} provides insight into the question of how a finite autocorrelation 
time can impact phase space transport in a complex phase space.

Generating white noise numerically is comparatively straightforward, 
requiring little more than producing a sequence of pseudo-random
impulses. Generating colored noise takes more thought. The 
algorithm exploited here was motivated by the recognition that, in the context 
of a stochastic differential equation, a white noise random process $X(t)$ 
can serve as a source to define a colored noise process $Y(t)$. As a concrete 
example, consider how Gaussian white noise can be used to implement a random 
process with  an autocorrelation function given by (5.2). 

Given one stochastic process, $X(t)$, one can define a second stochastic
process, $Y(t)$, implicitly as a solution to the stochastic differential
equation 
\begin{equation}
{d^{3}{ Y(t) }\over dt^{3}} + 3 {\alpha} {d^{2}{ Y(t) }\over dt^{2}} 
+3{\alpha}^2{d{ Y(t) }\over dt} + {\alpha}^3 Y(t) = X(t).
\end{equation}
Since the coefficients in eq. (5.5) are time-independent constants and $X(t)$ 
is stationary, it is clear that, if this equation be solved as an
initial value problem, at sufficiently late times
$Y(t)$ can also be considered stationary provided only, as is true, that the
dynamical system (5.5) is stable. 

By expressing $X(t)$ and $Y(t)$ in terms of their Fourier transforms, it is
easy to see that, neglecting the effects of nontrivial boundary 
conditions (e.g., choosing boundary conditions at $t_{0}\to -\infty$), the 
spectral densities $S_{X}({\omega})$ and 
$S_{Y}({\omega})$ for the two processes satisfy
\begin{equation}
S_{Y}( \omega ) = \frac{ S_{X}( \omega ) }{({\omega}^2 +{\alpha}^2)^3}.
\end{equation}
Assuming, however, that $X(t)$ corresponds to white noise, $S_{X}({\omega})
{\;}{\equiv}{\;}{\sigma}_{X}^{2}$ is a constant, so that the stochastic
process with spectral density $S_{Y}$ necessarily corresponds to colored noise.
Indeed, by performing an inverse Fourier transform it becomes evident that
$S_{Y}({\omega})$ corresponds to the stochastic process (5.2) with
\begin{equation}
$${\sigma}_{col}^{2}=3{\pi}{\sigma}_{X}^{2}/(8{\alpha}^{5}) . 
\end{equation}
A colored random process defined by the Langevin equation 
(2.5) or (2.8), with an autocorrelation function of the form (5.2), is 
equivalent mathematically to a collection of white noise processes. Solving 
(5.5) for $Y(t)$ yields the  colored input required to solve (2.5) or (2.8).

The one remaining question involves normalisations. To compare
different colored noises with each other or with an appropriately defined 
white noise limit, one must decide what should be meant by noise with
variable autocorrelation time but fixed amplitude. This was done here by 
considering sequences of random processes with different values of ${\alpha}$ 
and, for each ${\alpha}$, selecting ${\sigma}_{col}^{2}$ such that
\begin{equation}
\int_{-\infty}^{\infty} K_{\alpha}({\tau})d{\tau} =
\int_{-\infty}^{\infty} K_{white}({\tau})d{\tau} . 
\end{equation}
In other words, fixed amplitude but variable autocorrelation time was assumed
to correspond to different colored noises for which the time integral of the 
autocorrelation function assumes the same value. 
Noting that
\begin{equation}
\int_{-\infty}^{\infty} K_{white}({\tau})d{\tau}= 2{\Theta}{\eta}, 
\end{equation}
it follows that, for the stochastic process (5.2),
\begin{equation}
K({\tau})={3{\alpha}{\eta}{\Theta}\over 8}\exp(-{\alpha}|{\tau}|)\,
{\Bigl(}1+{\alpha}|{\tau}|+{{\alpha}^{2}\over 3}{\tau}^{2}{\Bigr)} .
\end{equation}
The Ornstein-Uhlenbeck process requires a normalisation
\begin{equation}
K({\tau})={\alpha}{\eta}{\Theta}\,\exp(-{\alpha}|{\tau}|).
\end{equation}

In the experiments modeling intrinsic noise, the colored noise was augmented 
by a friction ${\eta}$, and the friction and noise were related by a linear 
Fluctuation-Dissipation Theorem in terms of a temperature ${\Theta}$. In the
experiments modeling extrinsic noise, the friction vanished and ${\Theta}$ had 
no independent meaning as a temperature. Aside from ${\alpha}$, which fixes the
autocorrelation time, all that matters is the quantity ${\eta}{\Theta}$, which
sets the amplitude of the noise.

As for white noise, the evolution of an ensemble of noisy colored 
orbits is a two stage process. After an initial epoch without escapes, during 
which different members of the ensemble diverge exponentially, escapes turn on 
abruptly, with the first percent of the orbits escaping within an interval 
$T(0.01)$ much shorter than the time before the first escape. Interestingly, 
though, the second phase is often more complex than what is observed for white 
noise. Instead of evolving in a fashion that is well fit pointwise by an 
exponential decrease, the number remaining, $N(t)$ often exhibits ``plateaux'' 
and ``jumps'' with (almost) no, and especially large, decreases. (A detailed
examination of the data reveals that such irregularities can also arise for
white noise and periodic driving, but that are usually much less 
conspicuous in that case.) 

FIGS. 8 - 11 exhibit $N(t)$ generated for an initial condition where such 
irregularities are comparatively small. FIGS. 8 and 9 exhibit data generated
for different $4800$ orbit ensembles generated, respectively, for the 
stochastic processes (5.1) and (5.2), for the same initial condition as FIG. 
5, evolved with ${\Theta}=10.0$, ${\alpha}=0.2$, and variable ${\eta}$. 
FIGS. 10 and 11 exhibit analogous plots for the same initial condition with 
${\eta}=10^{-5}$ and variable ${\alpha}$. FIG. 12 exhibits data for a second
initial condition in the same potential with the same energy for which the 
early-times irregularities are especially conspicuous.

These irregularities imply that a pure exponential fit is often not justified.
Moreover, even when such a fit {\it is} justified, one finds that, for fixed
${\alpha}$, significantly different values of ${\eta}$ can yield very similar 
slopes\cite{30}.
In this sense, it is not accurate to state unambiguously that the escape rate
scales logarithmically with amplitude. However, what does remain true is that,
overall, escapes tend to happen more slowly in the presence of lower amplitude
perturbations, and that any systematic amplitude dependence is very weak, 
certainly much weaker than a simple power law ${\propto}{\;}{\eta}^{-p}$ with 
$p$ of order unity. Moreover, even though the observed escape rates exhibit
considerable irregularities, the one percent escape time $T(0.01)$ does not.
As for the case of white noise and periodic driving, $T(0.01)$ scales 
logarithmically in amplitude. Several examples are 
exhibited in FIGS. 13 (a) and (c), which plot $T(0.01)$ as a function of
${\rm log}\,{\eta}$ for two different initial conditions. In each case, the 
diamonds represent an Ornstein-Uhlenbeck process and the triangles the 
stochastic process (5.2). In panel (a) ${\alpha}=2.0$; in panel (c) 
${\alpha}=0.2$.


As for white noise, one also finds that, seemingly independent
of ${\alpha}$, the presence or absence of friction is largely irrelevant.
This is, e.g., evident from Table 1 which, for two values of
${\alpha}$, namely ${\alpha}=\infty$ (white noise) and ${\alpha}=0.02$ 
(autocorrelation time ${\tau}=50$), exhibits $T(0.01)$ as a function of 
$\log \eta$ both in the presence and the absence of friction.

Another obvious conclusion is that the efficacy of colored noise is a 
decreasing function of ${\alpha}$, the quantity that sets the autocorrelation
time $t_{c}$. When ${\alpha}$ is very large, so that $t_{c}$ is extremely 
short, color has virtually no effect. However, as ${\alpha}$ decreases and 
$t_{c}$ increases, $T(0.01)$ increases. In particular, for 
${\alpha}{\;}{\ll}{\;}1$, this 
corresponding to an autocorrelation time that is long compared to a 
characteristic orbital time scale, $T(0.01)$ is typically much longer 
than what is found in the white noise limit. This behaviour is evident from 
FIGS. 13 (b) and (d)  which exhibit $T(0.01)$ as a function of 
${\rm log}\,{\alpha}$ for fixed ${\eta}=10^{-5}$. As for FIGS. 13 (a) and (c), 
the diamonds and triangles represent, respectively, the stochastic processes 
(5.1) and (5.2). The horizontal dashed line represents the white noise value
towards which the data converge for ${\alpha}\to\infty$. The obvious inference
from this, and other, plots is that, the dependence of $T(0.01)$ on ${\alpha}$ 
or $t_{c}$ is again roughly logarithmic. This is reminiscent
of the fact that, as discussed in Section III, the efficacy of periodic driving
tends to scale logarithmically in the driving frequency.


Determining the overall efficacy of colored noise as a source of accelerated 
phase space transport thus involves an interplay between amplitude
and autocorrelation time, each of which, in the ``interesting'' regions of
parameter space contributes logarithmically to $T(0.01)$ and (modulo the
aforementioned caveats) ${\Lambda}$.

\section{DISCUSSION}
Just as for diffusion triggered by low amplitude periodic driving, the overall 
efficacy of noise-induced diffusion of ``sticky'' chaotic orbits scales
logarithmically in the amplitude of the perturbation. For both white and
colored noise, the one percent escape time $T(0.01)$ scales logarithmically
in the amplitude of the perturbation and, at least for white noise, so does
the initial escape rate. The details of the perturbation seem largely 
unimportant: The presence or absence of a friction term appears immaterial, 
and allowing for (at least some forms of) multiplicative noise also has 
comparatively minimal effects. For the case of white noise the only thing that 
seems to matter is the amplitude.

This logarithmic dependence on amplitude implies that what regulates the 
overall efficacy of the noise in inducing phase space transport is how fast 
perturbed noisy orbits diverge from the original unperturbed orbit. Escapes 
entail a two stage process, namely (i) an early interval during which noisy 
orbits diverge from the unperturbed orbit without breaching cantori and (ii) a 
later interval during which, in many cases, orbits escape seemingly at random 
in a fashion that samples a Poisson process. That $T(0.01)$ scales 
logarithmically in amplitude reflects the fact that escapes typically begin 
once ${\delta}r_{rms}$, the {\it rms} separation between perturbed and 
unperturbed orbits, approaches a critical value comparable to the size of the 
region in which the ``sticky'' orbits are originally stuck. That the escape
rate tends in many cases to exhibit at least a rough logarithmic dependence
reflects the fact that, even after the noisy orbits have spread out to sample 
a near-invariant population inside the confining cantori, random kicks can 
facilitate phase space transport by helping the orbits to ``find'' holes in 
the cantori. 

That escapes begin when ${\delta}r_{rms}$, rather than 
${\delta}E_{rms}$, approaches a critical value has an important implication
for how one ought to envision the escape process. Naively, it is not completely
obvious whether friction and noise facilitate phase space transport by 
``jiggling'' individual orbits or, since they change the orbital energy, 
by ``jiggling'' the effective phase space hypersurface in which the orbits 
move. The fact that ${\delta}r_{rms}$ sets the scale on which things happen 
demonstrates that, in point of fact, the former interpretation is more 
natural. 

Allowing for colored noise can significantly reduce the rate of phase space 
transport, but only when the autocorrelation time $t_{c}$ becomes comparable 
to, or larger than, a characteristic crossing time. If the noise is such that 
the spectral density $S({\omega})$ has only minimal power at frequencies 
comparable to the natural frequencies of the unperturbed orbit, its efficacy 
in inducing accelerated phase space transport will be reduced significantly. 
For ${\alpha}$ sufficiently large and $t_{c}$ sufficiently small, the two 
colored noises that were explored yield essentially the same results as did 
white noise; and similarly, for ${\alpha}$ sufficiently small and $t_{c}$ 
sufficiently long, the effects of the noise must become essentially negligible,
so that one recovers the behaviour observed for an unperturbed orbit. For
intermediate values, however, the escape statistics {\it do} depend on the 
value of ${\alpha}$. Moreover, this dependence is reminiscent of the effects
of periodic driving in at least one important respect: For the case of periodic
driving, quantities like $T(0.01)$ exhibit a roughly logarithmic dependence
on the driving frequency ${\omega}$. For the case of colored noise, $T(0.01)$
exhibits a roughly logarithmic dependence on ${\alpha}$. 

This suggests strongly that, like modulational diffusion triggered
by periodic driving, noise-induced phase space diffusion is intrinsically a
resonance phenomenon. If the Fourier transform of the noise has appreciable
power in the frequency range where the unperturbed orbit has appreciable power,
noise-induced diffusion will be comparatively efficient. If, however, the
noise has little power at such frequencies, it will serve as
a much less efficient agent for phase space transport, although the effects
need not be completely negligible.

To the extent that, as suggested by the numerical experiments described here,
the details are relatively unimportant, the effects of noise as a 
source of phase space transport are determined by two physical quantities, 
namely (i) the amplitude  and (ii) the autocorrelation time. Increasing the 
amplitude makes noise more important; increasing the autocorrelation time 
makes noise less important. 

As a concrete example, consider stars orbiting in an elliptical galaxy  
comparable in size to the Milky Way but located in the central part of a 
cluster like Coma, where the typical distance between galaxies is only five 
to ten times larger than the diameter of a typical galaxy. Here there are two 
obvious sources of noise which one might consider, namely (i) ``discreteness 
effects'' reflecting the fact that the galaxy is made of individual stars 
rather than a dustlike continuuum and (ii) the near-random influences of the 
surrounding environment. Discreteness effects result in gravitational 
Rutherford scattering, which can be modeled reasonably \cite{28} by 
friction and delta-correlated white noise related by a Fluctuation-Dissipation 
Theorem where, in natural units, ${\eta}{\;}{\sim}{\;}10^{-7}-10^{-9}$. To the
extent that the near random influences of the surrounding environment can be 
attributed primarily to a small number of neighbouring galaxies, it is also 
easy to estimate their amplitude and typical autocorrelation time.
Given that the nearest neighbouring galaxy is typically 
separated by a distance ${\sim}{\;}5-10$ times the diameter of the galaxy in 
question, and that the relative velocities of different galaxies in a cluster 
are usually comparable to the typical velocities of stars within an individual 
galaxy, one expects that the autocorrelation time $t_{*}$ is of order $5-10$ 
characteristic orbital times $t_{cr}$. Presuming, however, that the
perturbing influences of nearby galaxies reflect tidal effects, their overall
strength should scale as $D^{-3}$, where $D$ is the distance from the galaxy
in question, so that a typical amplitude ${\eta}{\;}{\sim}{\;}10^{-3}-10^{-2}$.

In this setting, discreteness effects give rise to comparatively weak noise 
with a very short autocorrelation time. Environmental effects given rise to a 
considerably stronger noise with a much longer autocorrelation time. The 
longer autocorrelation time tends to suppress the effects of environmental 
noise, but, even so, it would seem likely that, as a source of accelerated 
phase space transport, environmental noise will be significantly more 
important than discreteness noise.

\acknowledgments
The authors acknowledge useful discussions with Salman Habib and Katja
Lindenberg. Partial financial support was provided by the Institute for
Geophysics and Planetary Physics at Los Alamos National Laboratory. The
simulations involving colored noise were performed using computational
facilities provided by Los Alamos National Laboratory.

\vfill\eject
\onecolumn
\pagestyle{empty}
\begin{figure}[t]
\centering
\centerline{
        \epsfxsize=10cm
        \epsffile{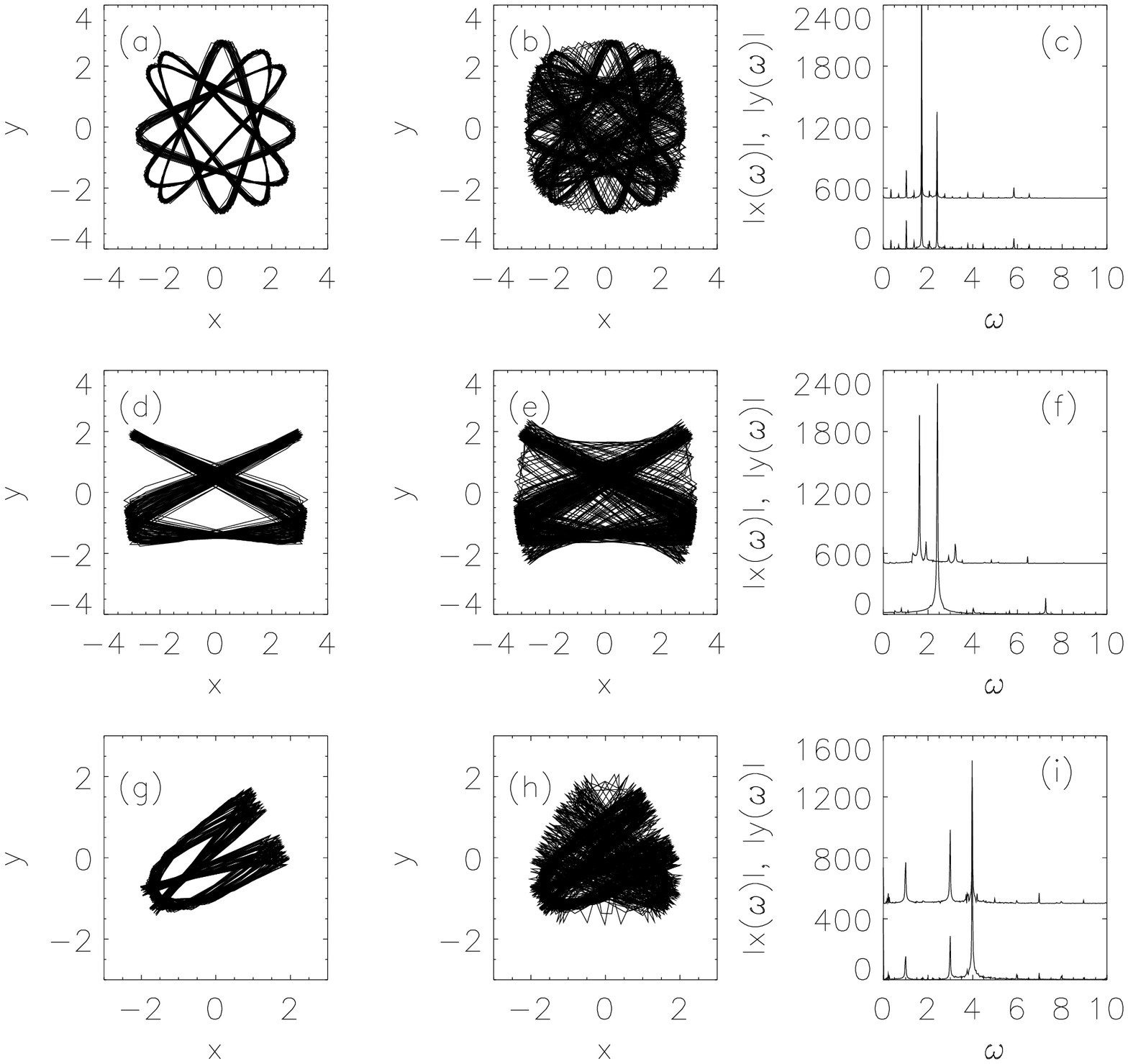}
           }
        \begin{minipage}{10cm}
        \end{minipage}
        \vskip -0.0in\hskip -0.0in
\caption{ (a) A chaotic initial condition with $E=10$ evolved in
the dihedral potential for a time $t=512$. (b) The same orbit integrated
for $t=1024$. (c) The power spectra $|x({\omega})|$ and $|y({\omega})|$ for
the orbit in (a).
(d) A chaotic initial condition with $E=20$ evolved in
the dihedral potential for a time $t=512$. (e) The same orbit integrated
for $t=1024$. (f) $|x({\omega})|$ and $|y({\omega})|$ for the orbit in (d).
(g) A chaotic initial condition with $E=20$ evolved in the
truncated Toda potential for a time $t=300$. (h) The same orbit integrated
for $t=600$. (i) $|x({\omega})|$ and $|y({\omega})|$ for the orbit in (g).}
\vspace{-5.0cm}
\end{figure}
\vfill\eject
\twocolumn
\pagestyle{empty}
\begin{figure}[t]
\centering
\centerline{
        \epsfxsize=8cm
        \epsffile{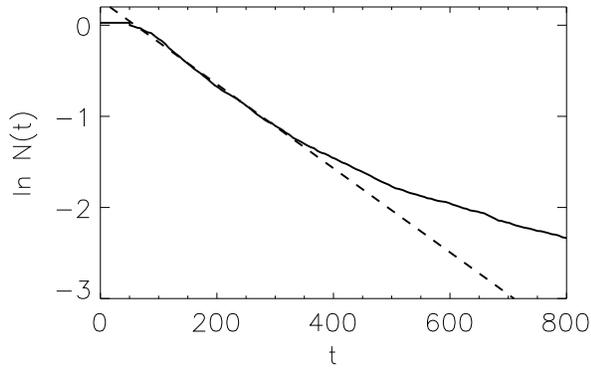}
           }
        \begin{minipage}{12cm}
        \end{minipage}
        \vskip -2.0in\hskip -0.0in
\caption{ $N(t)$, the fraction of the orbits from a $4001$ orbit ensemble not 
yet having escaped at time $t$,
computed for the initial condition exhibited in Fig. 1 (a), allowing for
a perturbation of amplitude $A=10^{-2.5}$ with variable frequencies 
$2.0{\;}{\le}{\;}{\omega}{\;}{\le}{\;}3.0$.}
\vspace{-0.2cm}
\end{figure}
\vfill\eject

\pagestyle{empty}
\begin{figure}[t]
\centering
\centerline{
        \epsfxsize=8cm
        \epsffile{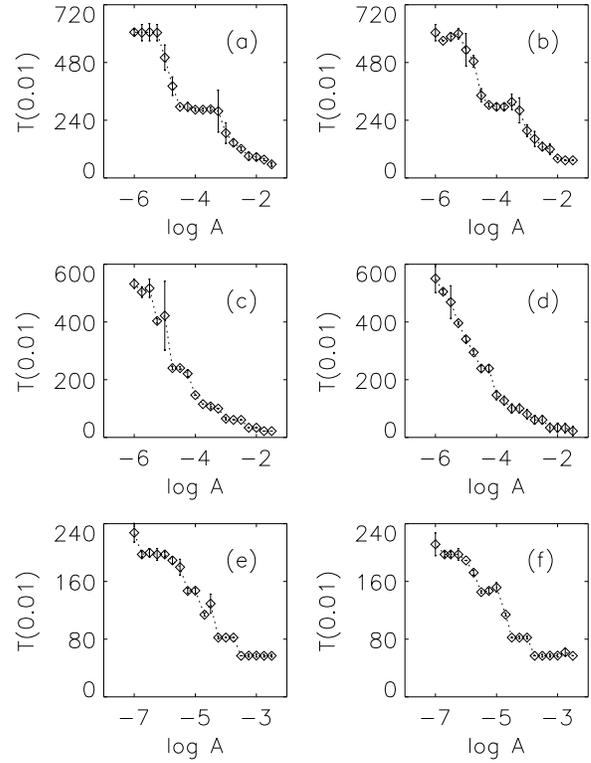}
           }
        \begin{minipage}{12cm}
        \end{minipage}
        \vskip 0.0in\hskip -0.0in
\caption{ (a) $T(0.01)$, the first escape time for 1\% of an ensemble of 
$1000$ 
integrations of the initial condition of Fig. 1 (a), driven with frequencies 
$0{\;}{\le}{\;}{\omega}{\;}{\le}{\;}1$, plotted as a function of the logarithm 
of the amplitude $A$ of the perturbation.
(b) The same for an ensemble with $3{\;}{\le}{\;}{\omega}{\;}{\le}{\;}4$.
(c) $T(0.01)$, the first escape time for 1\% of an ensemble of $1000$ 
integrations of the initial condition of Fig. 1 (d), driven with frequencies 
$0{\;}{\le}{\;}{\omega}{\;}{\le}{\;}1$, plotted as a function of the logarithm 
of the amplitude $A$ of the perturbation.
(d) The same for an ensemble with $3{\;}{\le}{\;}{\omega}{\;}{\le}{\;}4$.
(e) $T(0.01)$, the first escape time for 1\% of an ensemble of $1000$ 
integrations of the initial condition of Fig. 1 (g), driven with frequencies 
$0{\;}{\le}{\;}{\omega}{\;}{\le}{\;}1$, plotted as a function of the logarithm 
of the amplitude $A$ of the perturbation.
(f) The same for an ensemble with $3{\;}{\le}{\;}{\omega}{\;}{\le}{\;}4$.}
\vspace{-0.2cm}
\end{figure}
\vfill\eject

\pagestyle{empty}
\begin{figure}[t]
\centering
\centerline{
        \epsfxsize=8cm
        \epsffile{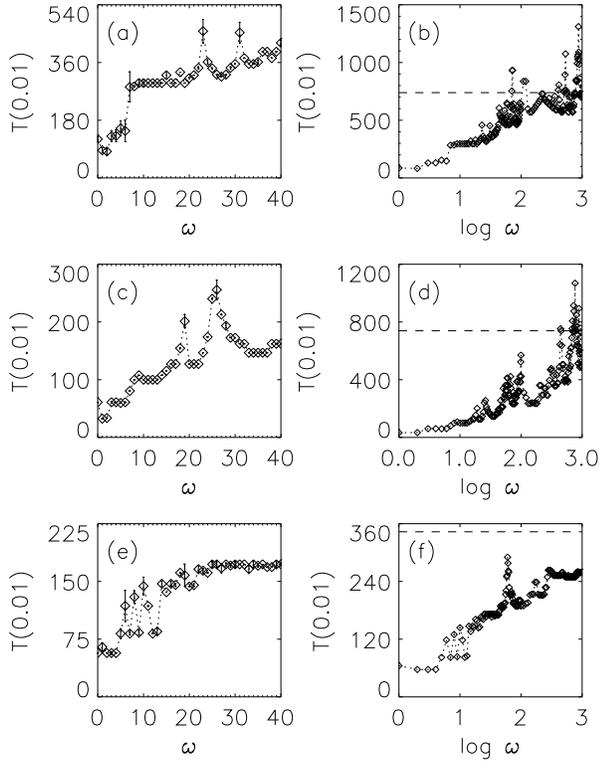}
           }
        \begin{minipage}{12cm}
        \end{minipage}
        \vskip 0.0in\hskip -0.0in
\caption{ (a) $T(0.01)$, the first escape time for 1\% of an ensemble of 
$1000$ 
integrations of the initial condition of Fig. 1 (a), driven with amplitude
$A=10^{-2.5}$, plotted as a function of frequency range 
${\omega}{\;}{\le}{\;}{\Omega}<{\omega}+1$ for $0<{\omega}<40$.
(b) The same information for $1<{\omega}<1000$, now plotted as a function of
$\log \omega$. The dashed line represents the escape time for the unperturbed
orbit. 
(c) $T(0.01)$, the first escape time for 1\% of an ensemble of $1000$ 
integrations of the initial condition of Fig. 1 (c), driven with amplitude
$A=10^{-2.5}$, plotted as a function of frequency range 
${\omega}{\;}{\le}{\;}{\Omega}<{\omega}+1$ for $0<{\omega}<40$.
(d) The same information for $1<{\omega}<1000$, now plotted as a function of
$\log \omega$. The dashed line represents the escape time for the unperturbed
orbit.
(e) $T(0.01)$, the first escape time for 1\% of an ensemble of $1000$ 
integrations of the initial condition of Fig. 1 (e), driven with amplitude
$A=10^{-2.5}$, plotted as a function of frequency range 
${\omega}{\;}{\le}{\;}{\Omega}<{\omega}+1$ for $0<{\omega}<40$.
(f) The same information for $1<{\omega}<1000$, now plotted as a function of
$\log \omega$. The dashed line represents the escape time for the unperturbed
orbit.}
\vspace{-0.2cm}
\end{figure}
\vfill\eject

\pagestyle{empty}
\begin{figure}[t]
\centering
\centerline{
        \epsfxsize=8cm
        \epsffile{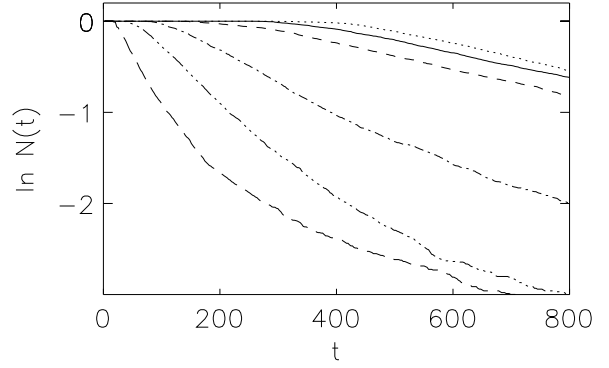}
           }
        \begin{minipage}{12cm}
        \end{minipage}
        \vskip -2.0in\hskip -0.0in
\caption{$N(t)$, the fraction of the orbits from a $2000$ orbit ensemble not 
yet having escaped at time $t$, computed for the initial condition exhibited 
in Fig. 1 (a), allowing for additive white noise with ${\Theta}=10$ and
variable ${\eta}=10^{-4}$ (broad dashes), ${\eta}=10^{-5}$ (triple-dot-dashed),
${\eta}=10^{-6}$ (dot-dashed), ${\eta}=10^{-7}$ (narrow dashes), 
${\eta}=10^{-8}$ (solid), and ${\eta}=10^{-9}$ (dots).}
\vspace{-0.2cm}
\end{figure}

\pagestyle{empty}
\begin{figure}[t]
\centering
\centerline{
        \epsfxsize=8cm
        \epsffile{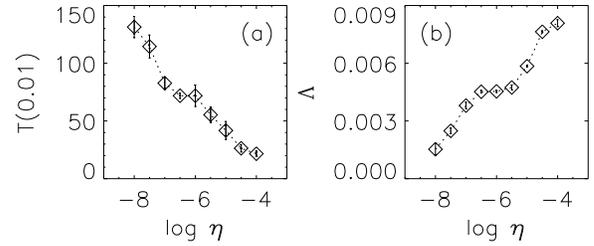}
           }
        \begin{minipage}{12cm}
        \end{minipage}
        \vskip -2.5in\hskip -0.0in
\caption{(a) $T(0.01)$, the first escape time for 1\% of an ensemble of $2000$ 
white noise integrations of the initial condition of Fig. 1 (a), with 
${\Theta}=10$ and variable ${\eta}$. (b) ${\Lambda}$, the rate at which orbits 
in this ensemble escape, fitted to the interval $T(0.01)<t<256$.}
\vspace{-0.2cm}
\end{figure}
\vfill\eject

\pagestyle{empty}
\begin{figure}[t]
\centering
\centerline{
        \epsfxsize=8cm
        \epsffile{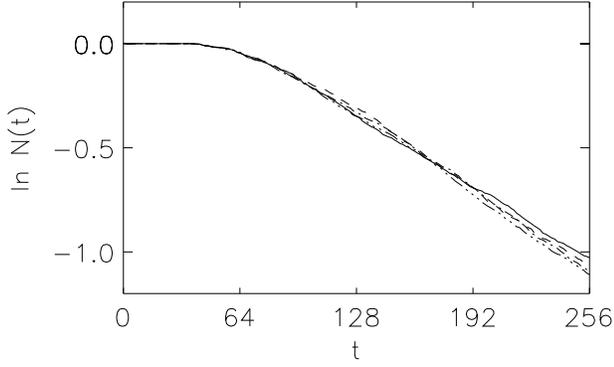}
           }
        \begin{minipage}{12cm}
        \end{minipage}
        \vskip -2.0in\hskip -0.0in
\caption{$N(t)$, the fraction of the orbits from a $2000$ orbit ensemble not 
yet having escaped at time $t$, computed for the initial condition exhibited 
in Fig. 1 (a) with ${\Theta}=10$ and ${\eta}_{0}=10^{-5}$. The four curves
represent additive white noise and constant ${\eta}$ (solid), additive 
white noise with no friction (dashed), multiplicative noise with
${\eta}{\;}{\propto}{\;}v^{2}$ (dot-dashed), and multiplicative noise with
${\eta}{\;}{\propto}{\;}v^{-2}$ (triple-dot-dashed).}
\vspace{-0.2cm}
\end{figure}

\pagestyle{empty}
\begin{figure}[t]
\centering
\centerline{
        \epsfxsize=8cm
        \epsffile{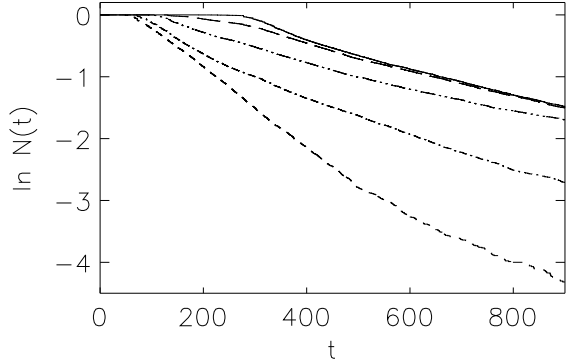}
           }
        \begin{minipage}{12cm}
        \end{minipage}
        \vskip -2.0in\hskip -0.0in
\caption{$N(t)$, the fraction of the orbits from a $4800$ orbit ensemble not 
yet having escaped at time $t$, computed for the initial condition exhibited 
in Fig. 1 (a), allowing for friction and additive Ornstein-Uhlenbeck noise 
with ${\alpha}=0.2$, ${\Theta}=10$, and variable ${\eta}=10^{-4}$ (dashed),
${\eta}=10^{-5}$ (dot-dashed), ${\eta}=10^{-6}$ (double-dot-dashed), 
${\eta}=10^{-7}$ (broad-dashed), and ${\eta}=10^{-8}$ (solid).}
\vspace{-0.2cm}
\end{figure}

\pagestyle{empty}
\begin{figure}[t]
\centering
\centerline{
        \epsfxsize=8cm
        \epsffile{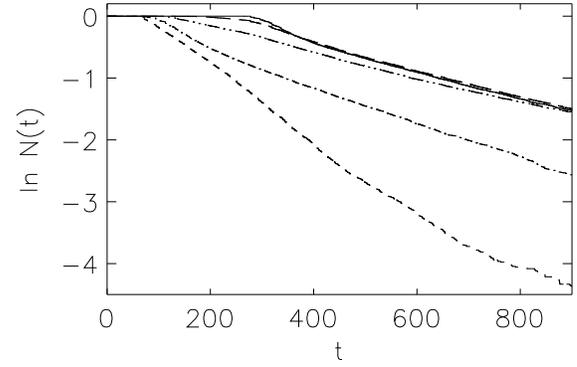}
           }
        \begin{minipage}{12cm}
        \end{minipage}
        \vskip -2.0in\hskip -0.0in
\caption{$N(t)$, the fraction of the orbits from a $4800$ orbit ensemble not 
yet having escaped at time $t$, computed for the initial condition exhibited 
in Fig. 1 (a), allowing for friction and colored noise given by eq. (5.2) 
with ${\alpha}=0.2$, ${\Theta}=10$, and variable ${\eta}=10^{-4}$ (dashed),
${\eta}=10^{-5}$ (dot-dashed), ${\eta}=10^{-6}$ (double-dot-dashed), 
${\eta}=10^{-7}$ (brad-dashed), and ${\eta}=10^{-8}$ (solid).}
\vspace{-0.2cm}
\end{figure}

\pagestyle{empty}
\begin{figure}[t]
\centering
\centerline{
        \epsfxsize=8cm
        \epsffile{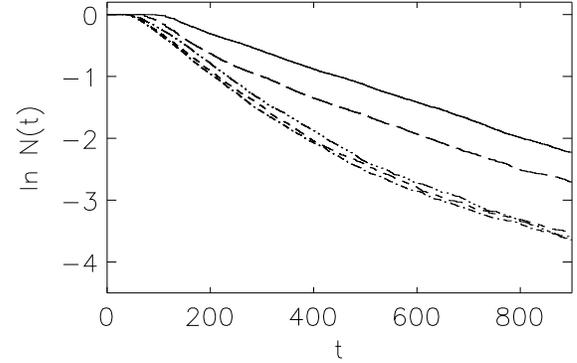}
           }
        \begin{minipage}{12cm}
        \end{minipage}
        \vskip -2.0in\hskip -0.0in
\caption{$N(t)$, the fraction of the orbits from a $4800$ orbit ensemble not 
yet having escaped at time $t$, computed for the initial condition exhibited 
in Fig. 1 (a), allowing for friction and Ornstein-Uhlenbeck noise, with
${\Theta}=10$, ${\eta}=10^{-5}$, and either white noise (dot-dashed) or 
variable ${\alpha}=20$ (dashed), ${\alpha}=2.0$ (triple-dot-dashed), 
${\alpha}=0.2$ (broad dashes), and ${\alpha}=0.02$ (solid)}
\vspace{-0.2cm}
\end{figure}

\pagestyle{empty}
\begin{figure}[t]
\centering
\centerline{
        \epsfxsize=8cm
        \epsffile{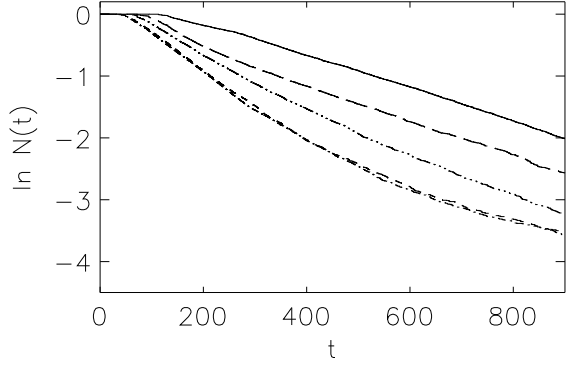}
           }
        \begin{minipage}{12cm}
        \end{minipage}
        \vskip -2.0in\hskip -0.0in
\caption{$N(t)$, the fraction of the orbits from a $4800$ orbit ensemble not 
yet having escaped at time $t$, computed for the initial condition exhibited 
in Fig. 1 (a), allowing for friction and colored noise given by eq. (5.2) with
${\Theta}=10$, ${\eta}=10^{-5}$, and either white noise (dashes) or variable 
${\alpha}=20$ (dot-dashed), ${\alpha}=2.0$ (triple-dot-dashed), ${\alpha}=0.2$ 
(broad dashes), and ${\alpha}=0.02$ (solid)}
\vspace{-0.2cm}
\end{figure}

\pagestyle{empty}
\begin{figure}[t]
\centering
\centerline{
        \epsfxsize=8cm
        \epsffile{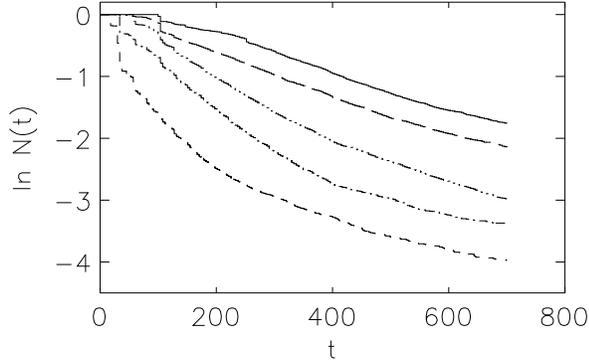}
           }
        \begin{minipage}{12cm}
        \end{minipage}
        \vskip -2.0in\hskip -0.0in
\caption{The same as FIG. 8 for a different initial condition.}
\vspace{-0.2cm}
\end{figure}

\pagestyle{empty}
\begin{figure}[t]
\centering
\centerline{
        \epsfxsize=8cm
        \epsffile{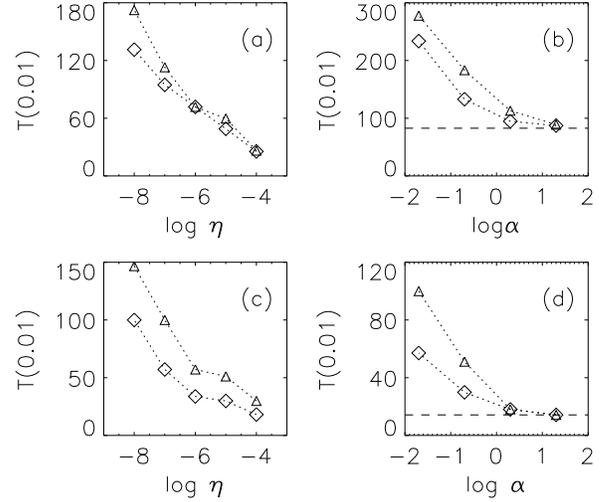}
           }
        \begin{minipage}{12cm}
        \end{minipage}
        \vskip -1.5in\hskip -0.0in
\caption{(a) $T(0.01)$, the first escape time for 1\% of an ensemble of $4800$
integrations, computed for the initial condition used to generate FIG. 1 (a),
plotted as a function of ${\rm log}\,{\eta}$ for fixed ${\alpha}=2.0$ for the 
stochastic processes defined by (5.1) (diamonds) and (5.2) (triangles), 
allowing for both friction and noise. (b) $T(0.01)$ for the same initial
condition, plotted as a function of ${\rm log}\,{\alpha}$ for fixed 
${\eta}=10^{-5}$, for the stochastic processes (5.1) (diamonds) and (5.2) 
(triangles), again allowing for both friction and noise. The dashed line
represents the asymptotic value for white noise $({\alpha}\to\infty)$.
(c) The same as (a), albeit for a different initial condition and with
${\alpha}=0.2$. (d) The same as (b), albeit for the initial condition in (c)
and with ${\eta}=10^{-7}$.
as (a) and (b) for another initial condition. }
\end{figure}

 \begin{table}

\caption{The 1\% escape time T(0.01) for orbits in an ensemble evolved in the
dihedral potential (2.1) with the initial condition used to generate Fig. 1 
(a), setting ${\Theta}=10.0$ and allowing for variable ${\eta}$. The colored
noise was generated by the stochastic process (5.2).
\label{table1}}
\vskip .1in
\begin{tabular}{rrrrr}
log ${\eta}$ & ${\alpha}={\infty}$ &${\alpha}={\infty}$
 & ${\alpha}=0.02$ & ${\alpha}=0.02$ \\ 
& with friction & no friction & with friction & no friction \\
&&&& \\
\tableline
&&&& \\
-4$\;\;\;$ & 21.621  &  21.672 &  71.930 &  71.811 \\
-5$\;\;\;$ & 41.740  &  41.660 & 113.024 & 112.980 \\
-6$\;\;\;$ & 71.863  &  71.816 & 172.396 & 172.393 \\
-7$\;\;\;$ & 82.744  &  82.724 & 277.684 & 277.682 \\
-8$\;\;\;$ & 131.368 & 131.616 & 277.704 & 277.709 \\
\end{tabular}
\end{table}


\begin{references}
\bibitem{30}This anomalous behaviour tends to arise most often for very small 
values of ${\eta}$, which suggests that, for sufficiently low amplitude, noise
is not all that helpful in allowing an ensemble which has already dispersed 
to find holes in the confining cantori.
\end{references}
\end{document}